\documentclass[preliminary]{eptcs}
\providecommand{\event}{WORDS 2011} 
\usepackage{breakurl}             

\usepackage[utf8]{inputenc}
\usepackage{tikz}
\usetikzlibrary{automata}
\usepackage{amssymb}
\usepackage{amsmath}
\usepackage{cancel}

\author{Thierry Monteil
\institute{CNRS -- Université Montpellier 2\\
\burl{http://www.lirmm.fr/~monteil}}
}

\def\titlerunning{The complexity of tangent words}
\def\authorrunning{Thierry Monteil}

\newtheorem{Theorem}{Theorem}

\newtheorem{proposition}[Theorem]{Proposition}

\title{The complexity of tangent words}

\date{}

\begin{document}

\newcommand{\qed}{\hfill $\square$}

\maketitle

\begin{abstract}
In \cite{MonteilDGCI2011}, we described the set of words that appear in
the coding of smooth (resp. analytic) curves at arbitrary small scale. The
aim of this paper is to compute the complexity of those languages.
\end{abstract}

\section{Introduction}

A \emph{smooth curve} is a map $\gamma$ from a compact interval $I$ of the
real line to the plane, which is $C^\infty$ and such that $||\gamma'(t)||
> 0$ for any $t\in I$ (this last property is called \emph{regularity}).
%
%
Any such curve can (and will be considered to) be arc-length
reparametrised (\emph{i.e.} $\forall t \in I, ||\gamma'(t)|| = 1$).
\newline
We can approximate such a curve by drawing a square grid of mesh $h$ on
the plane, and look at the sequence of squares that the curve meets.
For a generic position of the grid, the curve $\gamma$ does not hit any
corner and crosses the grid transversally, hence the curve passes from a
square to a square that is located either \emph{r}ight, \emph{u}p,
\emph{l}eft or \emph{d}own of it.
We record this sequence of moves and define the \emph{cutting sequence} of
the curve $\gamma$ with respect to this grid as a word $w$ on the alphabet 
$\{r,u,l,d\}$
%
which tracks the lines of the grid crossed by the curve $\gamma$.
\newline
The following picture shows a curve $\gamma$ with cutting sequence
$rruuldrrrd$.
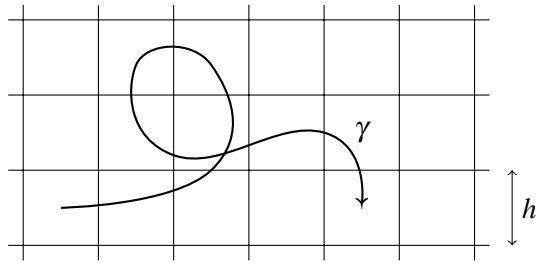
\begin{center}

\begin{tikzpicture}
    \draw[step=1,thin] (0,0) +(-0.2,-0.2) grid +(6.2,3.2);
    \draw[<->] (6.5,0) -- (6.5,1);
    \draw[right] (6.5,0.5) node {$h$};
    \draw[-<,thick] plot[smooth,tension=0.9] coordinates{(0.5,0.5) (2.5,1) (2.5,2.4) (1.5,2.4) (2,1.2) (4,1.5) (4.5,0.5) }; 
    \draw[thick] (4.5,1.5) node {\bf $\gamma$};
\end{tikzpicture}

\end{center}
Note that since the grid can be translated, a given curve may have more
than one cutting sequence for a given mesh $h$.
Our knowledge of the curve from one of its cutting sequences increases
when the mesh $h$ decreases, and when the mesh approaches $0$, the local
patterns of the cutting sequence play the role of discrete tangents. Such
words are called \emph{tangent words}, their first properties were
described in \cite{MonteilDGCI2011}.
Cutting sequences associated to straight segments are known to be exactly
the \emph{balanced words}, which are also the finite factors of Sturmian
words.
It turns out that the tangent words strictly contain balanced words, and
that $2$-balanced words strictly contain tangent words. The aim of this
note is to count the number of tangent words (resp. tangent analytic
words) of a given length, in order to quantify those inclusions.

%
%

\section{Tangent words}
Tangent words are the finite words that appear in the cutting sequences of
some smooth curve for arbitrary small scale.
More precisely, let $F(\gamma,G)$ denote the set of factors of the cutting
sequence of the curve $\gamma$ with respect to the square grid $G$ (when
the curve hits a corner, the cutting sequence is not defined and we set
$F(\gamma,G)=\emptyset$).
We define the \emph{asymptotic language} of $\gamma$ by
$$\displaystyle T(\gamma) = \limsup_{mesh(G) \rightarrow 0} F(\gamma,G) =
\bigcap_{\varepsilon > 0} \bigcup_{mesh(G) \leq \varepsilon} F(\gamma,G)
.$$
%
More generally, when $X$ is a set of curves, let us denote by $T(X)$ the
set $\bigcup_{\gamma\in X}T(\gamma)$. When $X$ is the set of smooth
curves, we denote $T(X)$ by $T^\infty$, and call its elements
\emph{tangent words}. When $X$ is the set of analytic curves, we denote
$T(X)$ by $T^\omega$, and call its elements \emph{analytic tangent words}.
The two languages $T^\infty$ and $T^\omega$ are factorial and extendable.
\newline \newline
For the sake of simplicity, we will focus on curves going right and up,
\emph{i.e.} smooth curves such that both coordinates of $\gamma'(t)$ are
positive for any $t$. Let us rename $r$ and $u$ by $0$ and $1$
respectively to stick to the usual notation about binary words.
\newline \newline
The following results are proved in \cite{MonteilDGCI2011}.

\subsection{Combinatorial characterisation (desubstitution)}

Balanced words are know to have a hierarchical structure, where the
morphisms $ \sigma_0 = ( 0 \mapsto 0 , 1 \mapsto 10 )$ and $ \sigma_1 = (
0 \mapsto 01 , 1 \mapsto 1 )$ play a crucial role \cite{PytheasFogg2002}
\cite{Lothaire2002}.
The same renormalisation applies to tangent words.
\newline
Given a finite word $w$, we can ``desubstitute'' it by 
\begin{itemize}
\item removing one $0$ per run of $0$ if $11$ does not appear in $w$, or
\item removing one $1$ per run of $1$ if $00$ does not appear in $w$.
\end{itemize}
This desubstitution map (denoted by $\delta$) consists in removing one
letter per run of the non-isolated letter. An accelerated version of this
desubstitution consists in removing a run equal to the length of the
shortest inner run from any run of the non-isolated letter (including
possible leading and trailing runs even if they have shorter length). 
\newline
If we repeat this process as much as possible, we get a \emph{derivated
word} denoted by $d(w)$. The word $w$ is balanced if, and only if, $d(w)$
is the empty word, and the derivation process is related to the continued
fraction development of the slope of the associated straight segment.

A word is said to be \emph{diagonal} if it is recognised by the following
automaton with three states, which are all considered as initial and
accepting: 
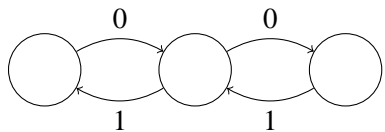
\begin{center}
\begin{tikzpicture} \node[state] (Q) {}; \node[state] (R) at (2,0) {};
\node[state] (S) at (4,0) {}; \path[->] (Q) edge [bend left]  node [above]
{$0$} (R); \path[->] (R) edge [bend left] node [below] {$1$} (Q);
\path[->] (R) edge [bend left]  node [above] {$0$} (S); \path[->] (S) edge
[bend left] node [below] {$1$} (R); \end{tikzpicture}
\end{center}

A word is said to be \emph{thin diagonal} if it is diagonal and only two
states are visited during its recognition.

A word is said to be \emph{non-oscillating diagonal} if it is recognised
by the following automaton with eight states, which are all considered as
initial and accepting:
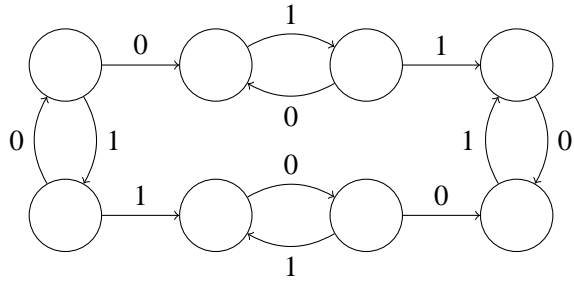
\begin{center}
\begin{tikzpicture}
    \node[state] (R) at (2,0) {};
    \node[state] (S) at (4,0) {};
    \node[state] (T) at (6,0) {};
    \node[state] (U) at (8,0) {};
%
    \node[state] (B) at (2,-2) {};
    \node[state] (C) at (4,-2) {};
    \node[state] (D) at (6,-2) {};
    \node[state] (E) at (8,-2) {};
    \path[->] (B) edge [bend left] node [left] {$0$} (R);
    \path[->] (R) edge [bend left] node [right] {$1$} (B);
    \path[->] (R) edge node [above] {$0$} (S);
    \path[->] (S) edge [bend left] node [above] {$1$} (T);
    \path[->] (T) edge [bend left] node [below] {$0$} (S);
    \path[->] (T) edge node [above] {$1$} (U);
    \path[->] (U) edge [bend left] node [right] {$0$} (E);
    \path[->] (E) edge [bend left] node [left] {$1$} (U);
%
    \path[->] (B) edge node [above] {$1$} (C);
    \path[->] (C) edge [bend left]  node [above] {$0$} (D);
    \path[->] (D) edge [bend left] node [below] {$1$} (C);
    \path[->] (D) edge node [above] {$0$} (E);
%
\end{tikzpicture}
\end{center}
\begin{proposition}
A finite word $w$ is tangent if, and only if, $d(w)$ is diagonal.
\newline
A finite word $w$ is tangent analytic if, and only if, $d(w)$ is
non-oscillating diagonal.\\
\end{proposition}
For example, the word $w = 100100010010010010001001000100$ is tangent
analytic since it can be desubstituted as
$1\cancel{00}1\cancel{00}01\cancel{00}1\cancel{00}1\cancel{00}1\cancel{00}01\cancel{00}1\cancel{00}01\cancel{00}
= 110111101101$, and then
$\cancel{11}0\cancel{11}110\cancel{11}0\cancel{1} = 01100 = d(w)$, which
is non-oscillating diagonal (start from the bottom left state).

\subsection{Geometric characterisation}
\begin{proposition}
A word $w$ is tangent if, and only if, for any $\varepsilon > 0$, $w$ is
the cutting sequence of a smooth curve $\gamma$ which is
$\varepsilon$-close (for the $C^1$ norm) to a straight segment (the grid
is fixed).
%
\newline
A word $w$ is tangent analytic if, and only if, for any $\varepsilon > 0$,
$w$ is the cutting sequence of a smooth curve $\gamma$ with nowhere zero
curvature which is $\varepsilon$-close (for the $C^1$ norm) to a straight
segment (the grid is fixed).\\
\end{proposition}
For example, the word $0110100110$ is tangent and the word $1001010110$ is
tangent analytic:
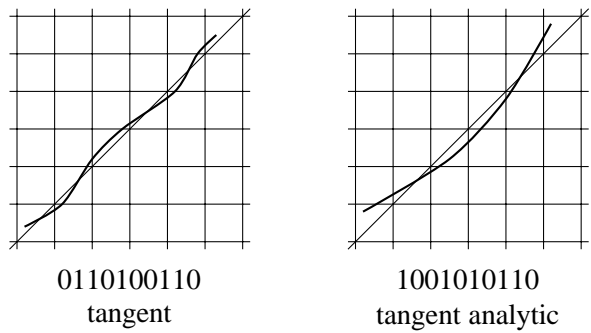
\begin{center}
\begin{tikzpicture}[scale=0.5]
\begin{scope}[shift={(-5,0)}]
    \draw[step=1,thin] (0,0) +(-0.2,-0.2) grid +(6.2,6.2);
    \draw (-0.2,-0.2) -- (6.2,6.2);
    \draw[thick] plot[smooth,tension=.5] coordinates{(0.2,0.4) (1.2,1) (2,2.2) (2.8,3) (4.2,4) (4.8,5) (5.3,5.5)}; 
    \draw (3,-1) node {$0110100110$};
    \draw (3,-2) node {tangent};
\end{scope}
\begin{scope}[shift={(4,0)}]
    \draw[step=1,thin] (0,0) +(-0.2,-0.2) grid +(6.2,6.2); 
    \draw (-0.2,-0.2) -- (6.2,6.2); 
    \draw[thick] plot[smooth,tension=.5] coordinates{(0.2,0.8) (2.5,2.2) (4,3.8) (5.2,5.8) };
    \draw (3,-1) node {$1001010110$};
    \draw (3,-2) node {tangent analytic};
\end{scope}
\end{tikzpicture}
\end{center}

\section{Complexity}

The \emph{complexity} of a language $L$ is the map that counts, for any
integer $n$, the number of elements of $L$ of length $n$. It is usually
denoted by $p_n(L)$.
\newline
The complexity of the balanced words $B$ was studied in
\cite{Lipatov1982}, \cite{Mignosi1991} and \cite{BerstelPocchiola1993},
where it was proved to be equal to:
$$ p_n(B) = 1 + \sum_{i=1}^{n}\sum_{j=1}^{i}\varphi(j) = 1 +
\sum_{i=1}^{n}(n-i+1)\varphi(i) \ ,$$
where $\varphi$ denotes the Euler totient function: $\varphi(n) = card \{
k \leq n \mid \gcd(k,n) = 1\}$.
\newline
\newline
To compute the complexity of $T^\infty$ and $T^\omega$, we will use the
tools introduced by Julien Cassaigne using bispecial factors
\cite{Cassaigne1997}. They have been used in the context of billiards in
\cite{CassaigneHubertTroubetzkoy2002}.
Let $L$ be a factorial and extendable language on the alphabet $\{0,1\}$.
A word $w$ in $L$ is said to be \emph{bispecial} if $0w$, $1w$, $w0$, $w1$
are in $L$. A bispecial factor $w$ is called 
\begin{itemize}
\item \emph{weak bispecial} if $card\{(a,b) \in \{0,1\}^2 \mid awb \in L \} = 2$, 
\item \emph{ordinary bispecial} if $card\{(a,b) \in \{0,1\}^2 \mid awb \in L \} = 3$,
\item \emph{strong bispecial} if $card\{(a,b) \in \{0,1\}^2 \mid awb \in L \} = 4$. 
\end{itemize}
Let $wb_n(L)$ (resp. $sb_n(L)$) denote the number of weak (resp. strong)
bispecial factors of length $n$ in $L$.
Let $s_n(L)$ denote the first difference $p_{n+1}(L) - p_n(L)$. We have:
$$s_{n+1}(L) - s_{n}(L) = sb_n(L) - wb_n(L) \ .$$
Hence, by summing twice, if $L$ is nontrivial, we have:
$$ p_n(L) = 1 + n + \sum_{i=0}^{n-1}\sum_{j=0}^{i-1} (sb_j(L) - wb_j(L)) \ .$$
\newline
Let us first describe the combinatorial structure of bispecial factors in
$T^\infty$.
Let $w$ be a bispecial factor.
If $w$ is not diagonal, then it can be desubstituted (in a single way) and
$\delta(w)$ is a bispecial factor of the same kind.
Otherwise, if $w$ is thin diagonal, then it is strong or ordinary
bispecial depending on the parity of its length. Otherwise, $w$ is
diagonal and the three states are visited during its recognition: $w$ is
strong bispecial.
Hence, there is no weak bispecial factor in $T^\infty$.
This also holds for $T^\omega$.
\newline
\newline
The geometric characterisation of tangent (resp. tangent analytic) words
is convenient to describe and count the strong bispecial factors.
We can visualise the strong bispecial factors as follows.
Pick a segment from  $(0,0)$ to $(p,q) \in \mathbb{Z}_{>0}^2$.
\newline
If there is no integer point on the way (which happens precisely when
$\gcd(p,q) = 1$), the coding of the corresponding open interval is a
bispecial factor of length $p+q-2$ in both $T^\infty$ and $T^\omega$.
Those words are also the bispecial factors for balanced words. There are
$\varphi(n+2)$ such words of length $n$, this the geometrical meaning of
Lipatov's formula \cite{Lipatov1982}.
\newline \newline

\begin{center}
\begin{tikzpicture}[scale=0.5]
\begin{scope}[shift={(-5,0)}]
\foreach \n in {10}{
    \foreach \var in {1,3,7,9}{\draw[above right] (\var,\n-\var) node {$1$} ;}

    \clip (-0.2,-0.2) -- (\n+0.5,-0.2) -- (-0.2,\n+0.5) -- cycle;
    \draw[step=1,thin] (0,0) +(-0.2,-0.2) grid +(\n+1,\n+1);
    \fill [black] (0,0) circle (4pt);

    \foreach \var in {1,...,\n}{
    \draw[very thin] (0,0) -- (\var,\n-\var);
    }

    \foreach \var in {1,3,7,9}{
    \draw[thick] (0,0) -- (\var,\n-\var);
    \fill [black] (\var,\n-\var) circle (4pt);
    }
}

\end{scope}
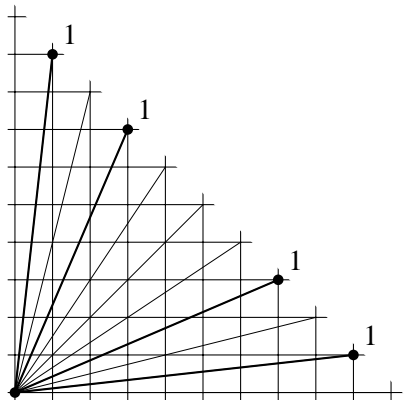
\draw (0,-1) node {Balanced bispecial factors of length $8$} ;
\end{tikzpicture}
\end{center}
Otherwise, there are $k\geq 1$ points one the way. 
For tangent analytic words, each such segment corresponds to two bispecial
factors of length $p+q-2$: one bending above the $k$ points, another bending under
the $k$ points. There are $2(n + 2 - \varphi(n+2))$ such words of length $n$.

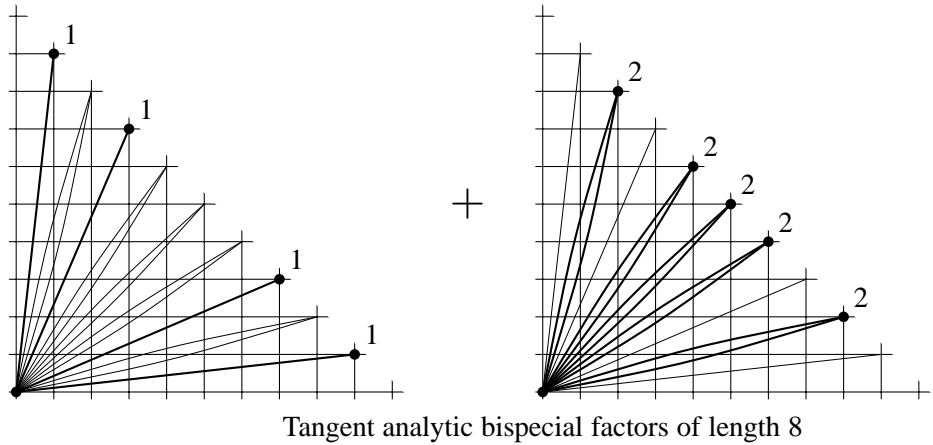
\begin{center}
\begin{tikzpicture}[scale=0.5]

\draw (-2,5) node {\LARGE $+$};

\begin{scope}[shift={(-14,0)}]
\foreach \n in {10}{
    \foreach \var in {1,3,7,9}{\draw[above right] (\var,\n-\var) node {$1$} ;}

    \clip (-0.2,-0.2) -- (\n+0.5,-0.2) -- (-0.2,\n+0.5) -- cycle;
    \draw[step=1,thin] (0,0) +(-0.2,-0.2) grid +(\n+1,\n+1);
    \fill [black] (0,0) circle (4pt);

    \foreach \var in {1,3,7,9}{
    \draw[thick] (0,0) -- (\var,\n-\var);
    \fill [black] (\var,\n-\var) circle (4pt);
    }


    \foreach \var in {2,4,5,6,8}{
    \draw[very thin] (0,0) .. controls (\var/2 - \n/40 + \var/40 , \n/2 - \var/2 + \var/40)  ..  (\var,\n-\var);
    \draw[very thin] (0,0) .. controls (\var/2 + \n/40  - \var/40 , \n/2 - \var/2 - \var/40)  ..  (\var,\n-\var);
    }

}
\end{scope}

\begin{scope}
\foreach \n in {10}{
    \foreach \var in {2,4,5,6,8}{\draw[above right] (\var,\n-\var) node {$2$} ;}

    \clip (-0.2,-0.2) -- (\n+0.5,-0.2) -- (-0.2,\n+0.5) -- cycle;
    \draw[step=1,thin] (0,0) +(-0.2,-0.2) grid +(\n+1,\n+1);
    \fill [black] (0,0) circle (4pt);

    \foreach \var in {1,3,7,9}{
    \draw[very thin] (0,0) -- (\var,\n-\var);
    }


    \foreach \var in {2,4,5,6,8}{
    \draw[thick] (0,0) .. controls (\var/2 - \n/40 + \var/40 , \n/2 - \var/2 + \var/40)  ..  (\var,\n-\var);
    \draw[thick] (0,0) .. controls (\var/2 + \n/40  - \var/40 , \n/2 - \var/2 - \var/40)  ..  (\var,\n-\var);
    \fill [black] (\var,\n-\var) circle (4pt);
    }

}
\end{scope}

\draw (0,-1) node {Tangent analytic bispecial factors of length $8$} ;
\end{tikzpicture}
\end{center}
For tangent words, each such segment corresponds to $2^k$ bispecial
factors of length $p+q-2$ corresponding to all the possibilities of
slaloming around the $k$ integer points on the way. 
Hence, there are $\displaystyle \sum_{\substack{d|n+2\\ d\neq 1}}
\varphi(n+2) 2^{(n+2)/d-1}$ strong bispecial factors of length $n$ in
$T^\infty$.

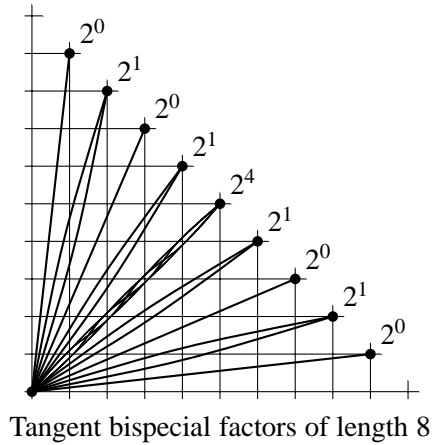
\begin{center}
\begin{tikzpicture}[scale=0.5]

\begin{scope}[shift={(-5,0)}]
\foreach \n in {10}{

    \foreach \var in {1,3,7,9}{\draw[above right] (\var,\n-\var) node {$2^0$} ;}
    \foreach \var in {2,4,6,8}{\draw[above right] (\var,\n-\var) node {$2^1$} ;}
    \draw[above right] (\n/2,\n/2) node {$2^4$} ;

    \clip (-0.2,-0.2) -- (\n+0.5,-0.2) -- (-0.2,\n+0.5) -- cycle;
    \draw[step=1,thin] (0,0) +(-0.2,-0.2) grid +(\n+1,\n+1);
    \fill [black] (0,0) circle (4pt);

    \foreach \var in {1,3,7,9}{
    \draw[thick] (0,0) -- (\var,\n-\var);
    \fill [black] (\var,\n-\var) circle (4pt);
    }


    \foreach \var in {2,4,6,8}{
    \draw[thick] (0,0) .. controls (\var/2 - \n/40 + \var/40 , \n/2 - \var/2 + \var/40)  ..  (\var,\n-\var);
    \draw[thick] (0,0) .. controls (\var/2 + \n/40  - \var/40 , \n/2 - \var/2 - \var/40)  ..  (\var,\n-\var);
    \fill [black] (\var,\n-\var) circle (4pt);
    }
    \foreach \i in {-1,1}{
        \foreach \j in {-1,1}{
            \foreach \k in {-1,1}{
                \foreach \l in {-1,1}{
                    \draw[-] plot[smooth,tension=0.9] coordinates{(0,0)
(1+\i/20,1-\i/20) (2+\j/20,2-\j/20) (3+\k/20,3-\k/20) (4+\l/20,4-\l/20)
(\n/2,\n/2)};

    }
    }        
    }
    }

    \fill [black] (\n/2,\n/2) circle (4pt);

}
\end{scope}

\draw (0,-1) node {Tangent bispecial factors of length $8$} ;
\end{tikzpicture}

\end{center}


\begin{proposition}
We have:

$$ p_n(T^\omega) = 1+n+\sum_{i=1}^{n}\sum_{j=2}^{i} (2j -\varphi(j) - 1)$$

$$ p_n(T^\infty) = 1+n+\frac{1}{2}\sum_{i=1}^{n}\sum_{j=2}^{i}\sum_{\substack{d|j\\d\neq 1}} \varphi(j) 2^{j/d} $$

\end{proposition}

\section{Conclusion}

Let us recall that a word $w$ is $k$-\emph{balanced} if:

$$\forall u,v \in Fact(w) \ \ \ \ |u| = |v| \Rightarrow | |u|_1 - |v|_1
|\leq k \ .$$

Each class of words is strictly included in the next one:
\begin{itemize}
\item $1$-balanced words (digital straight segments)
\item tangent analytic words
\item tangent words
\item $2$-balanced words \\
\end{itemize}
The complexity of the first two classes, is cubical whereas the complexity
of the last two classes is exponential.
It can be shown that analytic tangent words can be written as a
concatenation of two $1$-balanced words. What is the gap between tangent
words and $2$-balanced words ?

\bibliographystyle{eptcs}
\bibliography{biblio}

\begin{thebibliography}{1}
\providecommand{\bibitemdeclare}[2]{}
\providecommand{\urlprefix}{Available at }
\providecommand{\url}[1]{\texttt{#1}}
\providecommand{\href}[2]{\texttt{#2}}
\providecommand{\urlalt}[2]{\href{#1}{#2}}
\providecommand{\doi}[1]{doi:\urlalt{http://dx.doi.org/#1}{#1}}
\providecommand{\bibinfo}[2]{#2}

\bibitemdeclare{article}{BerstelPocchiola1993}
\bibitem{BerstelPocchiola1993}
\bibinfo{author}{Jean Berstel} \& \bibinfo{author}{Michel Pocchiola}
  (\bibinfo{year}{1993}): \emph{\bibinfo{title}{A geometric proof of the
  enumeration formula for {S}turmian words}}.
\newblock {\sl \bibinfo{journal}{Internat. J. Algebra Comput.}}
  \bibinfo{volume}{3}(\bibinfo{number}{3}), pp. \bibinfo{pages}{349--355},
  \doi{10.1142/S0218196793000238}.

\bibitemdeclare{article}{CassaigneHubertTroubetzkoy2002}
\bibitem{CassaigneHubertTroubetzkoy2002}
\bibinfo{author}{J.~Cassaigne}, \bibinfo{author}{P.~Hubert} \&
  \bibinfo{author}{S.~Troubetzkoy} (\bibinfo{year}{2002}):
  \emph{\bibinfo{title}{Complexity and growth for polygonal billiards}}.
\newblock {\sl \bibinfo{journal}{Ann. Inst. Fourier (Grenoble)}}
  \bibinfo{volume}{52}(\bibinfo{number}{3}), pp. \bibinfo{pages}{835--847}.
\newblock \urlprefix\url{http://aif.cedram.org/item?id=AIF_2002__52_3_835_0}.

\bibitemdeclare{article}{Cassaigne1997}
\bibitem{Cassaigne1997}
\bibinfo{author}{Julien Cassaigne} (\bibinfo{year}{1997}):
  \emph{\bibinfo{title}{Complexit\'e et facteurs sp\'eciaux}}.
\newblock {\sl \bibinfo{journal}{Bull. Belg. Math. Soc. Simon Stevin}}
  \bibinfo{volume}{4}(\bibinfo{number}{1}), pp. \bibinfo{pages}{67--88}.
\newblock
  \urlprefix\url{http://projecteuclid.org/getRecord?id=euclid.bbms/1105730624}.
\newblock \bibinfo{note}{Journ{\'e}es Montoises (Mons, 1994)}.

\bibitemdeclare{article}{Lipatov1982}
\bibitem{Lipatov1982}
\bibinfo{author}{E.~P. Lipatov} (\bibinfo{year}{1982}): \emph{\bibinfo{title}{A
  classification of binary collections and properties of homogeneity classes}}.
\newblock {\sl \bibinfo{journal}{Problemy Kibernet.}} (\bibinfo{number}{39}),
  pp. \bibinfo{pages}{67--84}.

\bibitemdeclare{book}{Lothaire2002}
\bibitem{Lothaire2002}
\bibinfo{author}{M.~Lothaire} (\bibinfo{year}{2002}):
  \emph{\bibinfo{title}{Algebraic combinatorics on words}}.
\newblock {\sl \bibinfo{series}{Encyclopedia of Mathematics and its
  Applications}}~\bibinfo{volume}{90}, \bibinfo{publisher}{Cambridge University
  Press}, \bibinfo{address}{Cambridge}.
\newblock \bibinfo{note}{Chapter 3, \emph{Sturmian Words} (by Jean Berstel and
  Patrice S\'e\'ebold)}.

\bibitemdeclare{article}{Mignosi1991}
\bibitem{Mignosi1991}
\bibinfo{author}{Filippo Mignosi} (\bibinfo{year}{1991}):
  \emph{\bibinfo{title}{On the number of factors of {S}turmian words}}.
\newblock {\sl \bibinfo{journal}{Theoret. Comput. Sci.}}
  \bibinfo{volume}{82}(\bibinfo{number}{1, Algorithms Automat. Complexity
  Games}), pp. \bibinfo{pages}{71--84}, \doi{10.1016/0304-3975(91)90172-X}.

\bibitemdeclare{inproceedings}{MonteilDGCI2011}
\bibitem{MonteilDGCI2011}
\bibinfo{author}{Thierry Monteil} (\bibinfo{year}{2011}):
  \emph{\bibinfo{title}{Another Definition for Digital Tangents}}.
\newblock In: {\sl \bibinfo{booktitle}{DGCI}}, {\sl \bibinfo{series}{Lecture
  Notes in Computer Science}} \bibinfo{volume}{6607}, pp.
  \bibinfo{pages}{95--103}, \doi{10.1007/978-3-642-19867-0\_8}.

\bibitemdeclare{book}{PytheasFogg2002}
\bibitem{PytheasFogg2002}
\bibinfo{author}{N.~Pytheas~Fogg} (\bibinfo{year}{2002}):
  \emph{\bibinfo{title}{Substitutions in dynamics, arithmetics and
  combinatorics}}.
\newblock {\sl \bibinfo{series}{Lecture Notes in Mathematics}}
  \bibinfo{volume}{1794}, \bibinfo{publisher}{Springer-Verlag},
  \bibinfo{address}{Berlin}, \doi{10.1007/b13861}.
\newblock \bibinfo{note}{Chapter 6, \emph{Sturmian Sequences} (by Pierre
  Arnoux)}.

\end{thebibliography}

\end{document}